\documentstyle[aps,pre]{revtex}
\newcommand{\gran}{\displaystyle}
\newcommand{\pert}{1.106\ldots}
\newcommand{\fitt}{1.08}

\newcommand{\asint}{0.2445}
\newcommand{\rsa}{0.74759\ldots}
\newcommand{\bd}{0.80865\ldots}
\begin{document}
\draft
\title{Irreversible Adsorption of particles after diffusing in a
gravitational field}
\author{Jordi Faraudo \thanks{E-mail:jordi@ulises.uab.es} and Javier
Bafaluy \thanks{E-mail:javier@ulises.uab.es}}
\address{Departament de F\'{\i}sica\\ Universitat Aut\`onoma de Barcelona\\
08193 Bellaterra (Barcelona), Spain}
\maketitle

\begin{abstract}
In this paper we analyze the influence of transport mechanisms
(diffusion and sedimentation) on the structure of monolayers of
particles irreversibly adsorbed on a line. We focus our attention on
the dependence of the radial distribution function $g(r)$ and the
saturation coverage $\theta_\infty$ on the gravitational P\'eclet
number $N_g$. First, we study the probability density of adsorption
onto an available interval using approximate solutions of the
transport equation and computer simulations. Combining our results
with an approximate general formalism, we can obtain values of
$\theta_\infty$ and the gap density, which agree with our
simulations. We also show that, for large gravity, the coverage
$\theta_{\infty}$ approaches the ballistic limit following a power
law in $N_g$ that is independent of the number of dimensions, as has
been observed in simulations \cite{Lleiescala}.
\end{abstract}

\section{Introduction}

The adsorption of particles of colloidal size (macromolecules,
latexes, bacteria, etc.) from fluid suspensions to solid surfaces is
a complex phenomenon of great interest. Much effort has been devoted
to the study of the mechanisms involved in this process
\cite{ADSORPTION}. A complete description should consider the
different steps involved: i) transport of the particles from the bulk
suspension toward the interface; ii) interaction with the substrate
(including the layer formed by the previously adsorbed particles);
iii) surface diffusion and desorption. For many colloidal particles,
for example some proteins \cite{FEDER,RAMSDEN} or latexes
\cite{ONODA,SCHAAF}, neither surface diffusion nor desorption is
observed in the time scale accessible to experiments: the particles
remain immobilized after adsorption, and the process can be
considered irreversible. Consequently, non-equilibrium configurations
are generated and, when the surface coverage attains a given value, a
{\em jamming} configuration is obtained, with no space available on
the surface for the adsorption of new particles.

The simplest model that attempts to describe these irreversible
adsorption phenomena, is the Random Sequential Adsorption (RSA) model
\cite {EVANS,RSA,RSAsint,RSAcomp,RSAg(r),RSAapr,Virial}. In this
model, particles are sequentially added to the surface by iteration
of the following algorithm: a) a random position is selected for the
addition of a new particle; b) if the new particle overlaps any
particle already adsorbed on the surface, the adsorption attempt is
rejected; c) if the new particle does not overlap any other particle,
then the adsorption attempt is accepted; d) the time necessary for
the adsorption of the particle is proportional to the number of
adsorption attempts. The RSA model has been extensively studied in
both discrete and continuous surfaces \cite{EVANS}. Exact results for
the kinetics, jamming coverage and distribution functions can be
obtained analytically for one-dimensional surfaces \cite{RSAg(r)};
for two-dimensional surfaces one has to use approximate methods
\cite{RSAapr,Virial} or computer simulations \cite {RSAcomp}.

Clearly, the RSA model does not consider the transport of the
particles toward the surface. Instead it assumes that new particles
arrive to the neighborhood of the surface with uniform probability,
and then they interact with the partially covered surface on the
basis of excluded volume interactions. More realistic models must
consider the different transport mechanisms that are present in
colloidal systems: diffusion, gravity, externally imposed flows, and
hydrodynamic and double layer forces \cite{COLLOIDS}.

For large enough particles suspended in a fluid at rest, gravity is
the dominant force: large particles sediment following a ballistic
trajectory toward the surface. This situation is described by the
ballistic deposition (BD) model \cite{BALLISTIC,BALexact}, in which
step b) of the RSA model is modified in the following way: if the
incoming particle overlaps a pre-adsorbed one, then it rolls down the
steepest descent path until either it reaches the surface and is
adsorbed, or it is trapped in a cavity over other particles, and is
rejected. This model is exactly solvable for one-dimensional surfaces
\cite{BALexact,BALgen}, and extensive approximate studies have been
realized for two-dimensional surfaces \cite{BALLISTIC}.

In the opposite limit of very small particles, Brownian motion
becomes the dominant transport mechanism. A model describing this
situation is the diffusion RSA model (DRSA) \cite{DRSA}: the initial
position of each particle is randomly chosen in a plane at a given
distance of the surface, and its trajectory is simulated by using a
Brownian dynamic algorithm \cite {ERMAK}; if the particle touches the
surface it adsorbs irreversibly. Furthermore, to avoid unbound
Brownian trajectories, particles arriving at points too far from the
surface are rejected. The simulation of this model is much more
expensive computationally than the RSA model and, furthermore, it
admits no exact solution even for one-dimensional surfaces.
Surprisingly, the jammed state obtained is very similar to the jammed
state obtained with the RSA model: very small differences appear in
one-dimensional surfaces, while in two-dimensional surfaces the
jamming states obtained from both models are indistinguishable with
the precision of the simulations. Analytically, good approximations
can be obtained for jammed one-dimensional surfaces from approximate
solutions of the diffusion equation \cite{DRSAB}.

For particles of general size both, the Brownian motion and gravity
have to be considered. Simulation studies have been done including
the gravity force in the Brownian dynamics algorithm \cite{DRSAG}.
The results show how the properties characterizing the surface
structure change smoothly with increasing gravity, from the DRSA
behavior in the limit of small gravity to the ballistic values in the
limit of large gravity. One interesting result is that the dependence
of the jamming coverage on the size of the particles seems to scale
to a common curve both for one and two-dimensional surfaces
\cite{Lleiescala}.  Therefore, the information obtained from 
one-dimensional surfaces can be relevant also to the more realistic
two-dimensional ones.

Finally, any realistic model has to include the effect of the
hydrodynamic interactions (HI), which become important when the particles
are in the neighborhood of the surface. Simulation results \cite{HYDRO,BDHYDRO}
suggest that the effect of HI on global averaged quantities
is small although they should be taken into account for a fine analysis
of the local structure. It can be useful therefore to study simpler models
that neglect hydrodynamic interactions but can be analytically solved.

Our aim in this paper is to develop an analytic approximation to the
description of the adsorption of particles on one-dimensional
surfaces considering the gravity force and Brownian motion and
neglecting hydrodynamic interactions. To do this, we first need
solutions for the transport equation in the presence of a flat
adsorbing surface and many particles fixed on it. These solutions can
be obtained with a sufficient approximation when only one particle is
fixed on the surface. Then, a superposition approximation has to be
done to study surfaces at finite coverage; this superposition has
shown good results for the DRSA model \cite{DRSAB}, and in the
present model it also gives good agreement with the simulations.

In section \ref{sec:description} we describe the model and the
analytical tools we use in detail. In section \ref{sec:oneparticle}
the one-particle effects are studied by approximate solution of the
transport equation and computer simulation. The results of this
section allow us to obtain an approximate description of the adsorbed
phase that is compared with simulations in
section\ref{sec:structure}. Finally, the appendices \ref{ap:multi}
and \ref{ap:layer} show how the approximate solutions of the
transport equation can be obtained.

\section{Description of the Model}
\label{sec:description}

We want to study a simple model that may allow us to understand the
effect on the structure of the adsorbed layer of two simple transport
mechanisms, namely diffusion and sedimentation. We consider an
adsorbing surface at $z=0$ and a semi-infinite fluid in the region
$z>0$. Spherical particles of radius $R$ suspended in the fluid
diffuse in the $XZ$ plane and sediment due to the effect of a uniform
gravitational field in the $Z$ direction. If the center of a new
particle arrives at the $z=0$ line, it is adsorbed irreversibly at
the contact point. We assume that the bulk concentration of particles
is so small that interactions between them are negligible.
Consequently, each particle adsorbs independently of the other
particles in the suspension, and the process can be considered as
{\em sequential}. However the concentration becomes large at the
surface, where adsorbed particles accumulate and interact via
excluded volume effects with incoming particles.

This adsorption problem can be studied in two steps: first, the
transport problem for a single particle near the surface in the
presence of the previously adsorbed particles has to be solved; then,
the evolution of the adsorbed phase can be studied from the obtained
adsorption probabilities.

Let $P(\vec r,t)$ be the probability density to find the
center-of-mass of a particle at the point $\vec r$ at time t; under 
the conditions of negligible inertia and small relaxation time, the
transport of the particle to the adsorbing line is governed by the
Smoluchowski equation,
\begin{equation}
\label{Smolu1}
\frac{\partial P(\vec{r},t)}{\partial t} =
\vec{\nabla}\cdot\left[D\vec{\nabla}P(\vec{r},t) +
uP(\vec{r},t)\hat{z}\right],
\end{equation}
$u$ being the sedimentation velocity and $D$ the diffusion
coefficient assumed constant. The probability flux is given by
\begin{equation}
\label{PFlux}
\vec{J}(\vec{r},t) = -D\vec{\nabla}P(\vec{r},t) -
uP(\vec{r},t)\hat{z}.
\end{equation}
The hard-particle interaction between bulk particles and adsorbed
ones is considered in the boundary conditions for eq. (\ref{Smolu1})
by assuming that the radial probability flux must vanish at the
exclusion surface (figure \ref{gap}) delimited by the preadsorbed
particles. At the adsorbing line we impose perfectly adsorbing
boundary conditions ($P=0$). Furthermore, we assume the initial
condition $P(\vec{r},t=0)=\delta(z-z_0)$, $z_0$ being the initial
distance of the center of the particle to the line; the value of
$z_0$ is not important provided that $z_0>2R$.

Equation (\ref{Smolu1}) with its boundary conditions can be made
dimensionless by measuring distances in units of the diameter $2R$ of
the particles and time in units of the characteristic diffusion time
$\tau_{diff}=4R^2/D$. Thus, the solution of (\ref{Smolu1}) depends on
a single dimensionless parameter, the gravitational P\'eclet number
$N_g$, defined as the quotient between $\tau_{diff}$ and the
characteristic sedimentation time $\tau_{det}=2R/u$,
\begin{equation}
\label{Peclet}
N_g=\frac{\tau_{diff}}{\tau_{det}}=\frac{2Ru}{D} = \frac{8\pi
R^4g\Delta\rho}{3k_BT},
\end{equation}
where $g$ is the acceleration of gravity, $\Delta\rho$ the difference
between the densities of the particles and the fluid, and $T$ the
absolute temperature. If $N_g\gg 1$ the motion of the particles is
deterministic and we recover the Ballistic Deposition model (denoted
in the following by BD), whereas if $N_g\ll 1$ Brownian motion
predominates and we recover the DRSA model \cite{DRSAB}. Note that
$N_g$ is proportional to $R^4$, and therefore we can define a
dimensionless radius as
\begin{equation}
\label{R*}
R^* \equiv R\left(\frac{4\pi g\Delta\rho}{3k_BT}\right)^{1/4} =
\left(N_g/2\right)^{1/4};
\end{equation}
the factor $2$ has been introduced to agree with the definition given
in \cite{DRSAG,Lleiescala}. For polystyrene beads in water at
$T=300K$, $\Delta \rho=45 kg/m^3$ and therefore $R^*\simeq 0.817 R$
if $R$ is expressed in microns. We will use these dimensionless
quantities in the following.

It is not possible to analytically solve the diffusion problem
(\ref{Smolu1}) in the general situation, when several adsorbed
particles are present. We will limit ourselves to obtentaining
approximate solutions in the presence of only one adsorbed particle.
To describe the general situation, then, we will introduce a
superposition approximation by assuming that the adsorption
probability at a given point depends only on the position of the
nearest particles, that is, those limiting the free interval of the
line to which the point belongs. One expects this hypothesis to fail
only for small gaps (and therefore only slightly affecting the layer
structure) if gravity is not too high. If gravity is strong as
compared to diffusion, when an incoming particle touches an
adsorbed one, it is expected to adsorb in its immediate vicinity and
the superposition hypothesis holds also for the smallest gaps.

According to the previous discussion, the solution of the transport
equation will lead to an adsorption model in which particles are
sequentially deposited onto an infinite line with an adsorption
probability density that depends on the gap in which adsorption takes
place. This means that the adsorbing particle only significantly
interacts with the particles that are limiting this gap, and
interactions with other particles are not considered. The solution of
the Smoluchowski equation (\ref{Smolu1}) gives the adsorption
probability at any point of an available interval. The problem of
characterizing the structure of the adsorbed monolayer can be studied
starting from the kinetic equation for the evolution of the free
gaps. We summarize the method below, and a more detailed discussion
can be found in \cite{DRSAB}.

Let $G(h,t)$ be the number density of gaps with length $h$ at time
$t$ and $k(h',h)$ the probability per unit length and per unit time
that the deposition of a particle in a gap of length $h'>1$ produces
gaps of length $h$ and $h'-h-1$. The governing kinetic equation for
the irreversible adsorption process is \cite{DRSAB} (see figure
\ref{gap}):
\begin{equation}
\label{bal}
\frac{\partial G(h,t)}{\partial t} = -k_0(h)G(h,t) +
2\int_{h+1}^\infty dh' G(h',t)k(h',h),
\end{equation}
where $k_0(h)$ is the total rate at which gaps of length $h$ are
destroyed by the addition of a new particle:
\begin{equation}
\label{defk0}
k_0(h)=\int_0^{h-1}dh' k(h,h').
\end{equation}
The gap distribution in the jamming state can be obtained without
knowing the detailed time evolution. From the balance equation
(\ref{bal}) it is possible to derive a time-independent equation for
the total number density of gaps of length $h$ that have been created
at any time along the process, $n(h)$ \cite{DRSAB}:
\begin{equation}
\label{int}
n(h)=2\int_{h+1}^\infty dh' p(h',h)n(h'),
\end{equation}
where $p(h',h)$ is the probability density that the first particle
arriving at an interval of length $h'$ creates two new free intervals
of length $h$ and $h'-h-1$,
\begin{equation}
\label{defp}
p(h',h) = \frac{k(h',h)}{k_0(h')}.
\end{equation}
From the definition of $k_0(h)$, eq. (\ref{defk0}), this function
satisfies the normalization
\begin{equation}
\label{normaP}
\int_0^{h'-1}p(h',h)dh=1.
\end{equation}
In addition eq. (\ref{int}) must be supplemented by a normalization
condition for $n(h)$. From the definition of $n(h)$ we have
\cite{DRSAB}
\begin{equation}
\label{norman}
\int_0^1(1+h)n(h)dh=1.
\end{equation}

A global measure of the particle packing is the coverage defined as
the relative length (area in $2$ dimensions) of a line of total
length $L$ covered by $N$ particles of radius $R$:
\begin{equation}
\theta =\frac{2RN}L.
\end{equation}
The coverage increases monotonically (due to the irreversible nature
of the process) until a saturation value $\theta_\infty$, when no
available space remains for adsorption of new particles. In this
situation only gaps with $h\le 1$ survive and the saturation coverage
can be obtained from $n(h)$ using the fact that the particle number
is equal to the number of gaps,
\begin{equation}
\label{rec}
\theta_\infty =\int_0^1n(h)dh.
\end{equation}
Let us remark that eq. (\ref{int}) expresses the fact that the total
number of gaps of length $h$ can be computed from the number of gaps
with length $h' > h+1$ and the probability $p(h',h)$ of obtaining an
interval of length $h$ from an interval of length $h'$. In principle,
from eqs. (\ref{int})-(\ref{rec}) it is possible to obtain $n(h)$ and
the saturation coverage $\theta_\infty $ if $p(h',h)$ is known. For
each specific adsorption model $p(h',h)$ must be previously known.

In order to obtain $p(h',h)$, we must solve eq. (\ref{Smolu1}) with
two adsorbed particles limiting a gap of length $h'$ according to the
hypothesis assumed in the derivation of eqs. (\ref{bal})-(\ref{int}).
The accuracy of this ansatz is analyzed by means of Brownian dynamics
simulations in section \ref{sec:structure}.

\section{Adsorption Probability}
\label{sec:oneparticle}

In this section, we study the probability density of the adsorption
of a new particle on a line in which one adsorbed sphere is already
present. To find this probability we compute approximate solutions of
the Smoluchowski equation (\ref{Smolu1}) and we perform Brownian
dynamics simulations. The results obtained will be used in the next
section to determine the structure and coverage of jammed lines.

To use the integral equation (\ref{int}), we need the probability
density $p(h',h)$ defined in (\ref{defp}), which can in principle be
obtained by solving the Smoluchowski equation (\ref{Smolu1}) in
presence of two preadsorbed particles separated by a center to center
distance $h'+1$. However, the solution to this problem is very
difficult and therefore we have adopted a superposition hypothesis
assuming that the joint effect of the two preadsorbed particles can
be obtained from a superposition of the one-particle effects.

The one-particle adsorption problem is defined as follows: the center
of an adsorbed particle is fixed at the point $(x,z)=(0,0)$ and, at
time $t=0$, a new particle starts to move from an initial point at a
given distance from the surface $z_0$, with a random initial value of
$x$. The initial height $z_0$ is immaterial and it will be assumed to
be infinite. In the presence of a non-vanishing gravity field, the
particle will be adsorbed with probability one. We normalize
$P(\vec{r}, t)$ in such a way that the initial probability per unit
length is equal to one. Therefore, the probability density of
adsorption at a point of the adsorbing line $z=0$ is:
\begin{equation}
\label{ro}
\rho(x)=-\int_0^\infty J_z(x,z=0,t)dt,
\end{equation}
where $J_z(x,z,t)$ is the $z$ component of the probability flux
defined in eq. (\ref{PFlux}). Using the superposition hypothesis, we
write for the probability density of adsorption on a gap of length
$h'$:
\begin{equation}
\label{superposicio}
p(h',h)\simeq N(h')\rho (h+1)\rho(h'-h),
\end{equation}
where $N(h')$ is determined by the normalization condition
(\ref{normaP}). To obtain $\rho(x)$ we do not need to solve the
time-dependent Smoluchowski equation (\ref{Smolu1}). We only need the
time-integrated probability density,
\begin{equation}
\label{fi}
\psi (\vec r)=\int_0^\infty dtP(\vec r,t), 
\end{equation}
that obeys the equation (obtained by the integration of
(\ref{Smolu1}) and using dimensionless units)
\begin{equation}
\label{Smolu2}
\nabla^2\psi (\vec r)+N_g\frac{\partial \psi (\vec r)}{\partial z}=0.
\end{equation}
The boundary conditions satisfied by $\psi$ are the perfect adsorbing
boundary condition on the line z=0:
\begin{equation}
\label{contorn1}
\psi(x,z=0)=0,
\end{equation}
and the condition that the total probability flux coming from
$z=\infty$ is $1$ per unit length,
\begin{equation}
\label{contorn1p}
\frac{\partial\psi(\vec r)}{\partial z} + N_g\psi(\vec r) \to 1,
\;\;\; z\to\infty.
\end{equation}

Using eq. (\ref{ro}) and the boundary condition (\ref{contorn1}),
$\rho (x)$ can be obtained from $\psi$ as
\begin{equation}
\label{ro2}
\rho (x)=\left. \frac{\partial \psi }{\partial z}\right|_{z=0.}
\end{equation}
Another boundary condition is necessary to account for the presence
of the previously adsorbed particle in the origin of coordinates. The
hard sphere interaction between diffusing and adsorbed particles
implies no radial flux at the exclusion sphere (of radius 2R)
centered at the origin; therefore we have (see figure \ref{gap}):
\begin{equation}
\label{contorn2}
\left(\frac{\partial\psi(r,\theta)}{\partial r}\right)_{r=1} +
N_g\psi(r=1,\theta )\sin\theta = 0.
\end{equation}
The solution $\psi (\vec r)$ can be split into two parts, 
\begin{equation}
\label{dosparts}
\psi (\vec r)=\psi^{(0)}(\vec r)+\psi^{(1)}(\vec r),
\end{equation}
where $\psi^{(0)}$ is the solution corresponding to a clean
adsorption line, and $\psi^{(1)}$ reflects the presence of the
adsorbed particle and must vanish at large distances. The solution of
(\ref{Smolu2}) for $\psi^{(0)}(\vec r)$ with conditions
(\ref{contorn1},\ref{contorn1p}) is
\begin{equation}
\label{fi0}
\psi^{(0)}(x,z)=\frac 1{N_g}\left( 1-e^{-N_gz}\right).
\end{equation}
The probability density of adsorption in the presence of the fixed
particle is obtained using eq. (\ref{ro2}):
\begin{equation}
\label{ro3}
\rho (x) = \left.\frac{\partial\psi^{(0)}}{\partial z}\right|_{z=0} +
\left.\frac{\partial\psi^{(1)}}{\partial z}\right|_{z=0} \equiv 1 +
\rho^{(1)},
\end{equation}
where $\rho^{(1)}$ is the deviation of $\rho$ from the uniform
distribution. In the interior of the region excluded by the fixed
particle $(|x|<1)$, the total flux has to be zero and therefore
$\rho^{(1)}=-1$. In the available region $(|x|>1)$, $\rho^{(1)}$
gives the local increase of the adsorption probability due to the
particles that have been in contact with the adsorbed particle; these
particles are rejected in the RSA model, but here are allowed to
diffuse and will adsorb in the neighborhood of the fixed particle.
The presence of the fixed particle does not change the total
adsorption probability, so the integral of $\rho^{(1)}$ over the
entire line has to vanish. Therefore one has
\begin{equation}
\label{normaro}
\int_1^\infty \rho^{(1)}dx=1.
\end{equation}

The equation and boundary conditions for $\psi^{(1)}(\vec r)$ can be
obtained by substitution of (\ref{dosparts}),(\ref{fi0}) in
(\ref{Smolu2}),(\ref{contorn1}),(\ref{contorn2}). It is not possible
to obtain a solution for general values of $N_g$, and different
approaches have to be used in the limits of small and large gravity.

\subsection{Small gravity}

We obtain a multipolar expansion for $\rho(x)$ by solving the
equation for $\psi^{(1)}(\vec r)$ and substitution in (\ref{ro3}). To
obtain $\psi^{(1)}(\vec r)$ we use the Green function method, for the
detailed calculations see Appendix \ref{ap:multi}. The (formally
exact) result for $\rho(x)$ can be expressed in the form
\begin{equation}
\label{multipolar}
\rho(x) = 1 + \frac 2{\pi |x|} \sum_{n=1}^\infty
(2n-1)\varphi_{2n-1}\frac{K_{2n-1}(\frac{N_g}{2}|x|)}
{K_{2n-1}(\frac{N_g}{2})},\;\;\;|x|>1,
\end{equation}
where $K_n(x)$ is the Bessel function of third kind and the
coefficients $\varphi_{2n-1}$ satisfy the infinite linear system of
equations (\ref{Ap6},\ref{Ap61}).

By substitution of the multipolar expression (\ref{multipolar}) in
(\ref{superposicio}) we obtain the probability density of adsorption
on a gap of length $h'$ for given $N_g$. The normalization factor
$N(h')$ is obtained by numerical integration of (\ref{normaP}). For
practical uses, the series (\ref{multipolar}) must be truncated by
using a finite number of terms; the larger the gravitational P\'eclet
number $N_g$ is, the larger the number of terms that have to be
retained in (\ref{multipolar}) to achieve a given precision. We
recall that if we want to obtain the saturation coverage
$\theta_\infty $ we have to solve the integral equation (\ref{int})
with the kernel $p(h',h)$ given by eqs (\ref{superposicio}),(\ref
{multipolar}). If $N_g$ is of order $1$ we need only a few terms in
the summation and the numerical solution of the integral equation is
possible, but for intermediate to high values of $N_g$ we need a very
large number of terms and this expansion becomes useless. Therefore,
it is necessary to obtain an asymptotic expression for large values
of $N_g$ to enable a description in the entire range of variation of
$N_g$. We will return to this question in the next subsection.

We can see how the DRSA result \cite{DRSAB} can be recovered by
taking the $N_g\to 0$ limit in the previous expression. For small
$N_g$ we need only one term in (\ref{multipolar}), because
$\varphi_{2n-1} = \frac{\pi}{2}\delta_{1n} + {\cal O}(N_g)$.
Therefore,
\begin{equation}
\label{ro0}
\rho(x) = 1 + \frac{1}{|x|} \frac{K_1(\frac{N_g}{2}|x|)}
{K_1(\frac{N_g}{2})} + {\cal O}(N_g) \approx 1 + 1/x^2.
\end{equation}
The last approximation has been obtained using the small argument
expansion of the Bessel function $K_1(x)\simeq \frac{1}{x}$. It gives
the known zero-gravity result \cite{DRSAB} and we see now that it
gives an approximation for small gravity valid in the region $x\ll
2/N_g$.

The first moment of $\rho^{(1)}$ gives the mean distance to the
origin where particles touching the fixed one adsorb. From eq.
(\ref{multipolar}), and using the well-known formula for the
derivatives of Bessel functions, $K_{n-1}(y) + K_{n+1}(y) =
-2K'_n(y)$ we obtain:
\begin{equation}
<|x|>_1 = \int_1^\infty \rho^{(1)}(x) x dx = \frac{2}{\pi}
\sum_{n=1}^\infty 
\frac{(2n-1)\varphi_{2n-1}(N_g)}{\frac{N_g}{2}K_{2n-1}(\frac{N_g}{2})}
(-1)^{n+1}\left[ K_0(\frac{N_g}{2}) +
2\sum_{i=2}^n(-1)^{i-1}K_{2i-2}(\frac{N_g}{2})\right].
\end{equation}
For $N_g\to 0$ only the first term of this sum has to be retained,
obtaining
\begin{equation}
<|x|>_1 \approx \frac{2}{N_g} \frac{K_0(N_g/2)}{K_1(N_g/2)}
\approx \log\left(\frac{4}{N_g}\right).
\end{equation}
The mean distance increases without bound, as can be expected from
the DRSA limit given by eq. (\ref{ro0}).

\subsection{Large gravity}

For large values of $N_g$ one can try to solve eq. (\ref{Smolu2}) by
a perturbative approach. In the $N_g\to\infty$ limit the higher order
derivatives in eq. (\ref{Smolu2}) are negligible, and an approximate
equation is obtained retaining only the advective term. This is
equivalent to completely neglecting diffusion. The corresponding
solution satisfying the boundary condition at infinity is $\psi =
1/N_g$, which does not verify the boundary conditions at $z=0$ and
$r=1$. The reason is that, no matter how large $N_g$ is, diffusion
becomes important near the boundaries of the system: $\psi$ must
change appreciably in a thin {\em boundary layer}, and the
corresponding derivatives are no longer negligible in that layer.

The problem can be solved using singular perturbation techniques as
discussed in appendix \ref{ap:layer}. The result is that, for large
values of $N_g$, $\rho^{(1)}(x)$ scales to a function of the form
\begin{equation}
\label{scaling}
\rho^{(1)}(x) \approx N_g^{2/3} \phi^{(1)}[(x-1)N_g^{2/3}],
\end{equation}
where function $\phi^{(1)}$ is defined in the appendix. The mean
distance at which a particle that hits the preadsorbed one can be
adsorbed is therefore a quantity of order $N_g^{-2/3}$,
\begin{eqnarray}
\label{distanciab}
<|x|>_1 &=& \int_1^\infty \rho^{(1)}(x)xdx \approx 1 +
\frac{A}{N_g^{2/3}}, \nonumber\\
A &=& \int_0^\infty \phi^{(1)}(\xi)\xi d\xi \approx \pert.
\end{eqnarray}
Obviously this distance decreases when $N_g$ grows because diffusion
is weak and the particles adsorb closer. If $N_g=\infty$, a particle
hitting the fixed sphere will roll over it and be adsorbed in its
immediate vicinity (at $|x|=1$) as corresponds to the rolling
mechanism of BD. If $N_g$ is large but finite, diffusion disturbs the
deterministic motion and the new particle can be adsorbed at some
distance from the adsorbed one, leaving between them a gap of size
$h\sim N_g^{-2/3}$.

We can give a simple reasoning supporting the scaling given by eq.
(\ref{distanciab}) for large values of $N_g$. Particles falling over
the fixed particle will roll down over it and diffuse away along a
thin layer of thickness $\delta$. This thickness can be estimated
from the condition that the radial probability flux due to gravity,
which is proportional to $N_g\sin\theta$, has to be compensated with
the diffusion flux that can be estimated as $\delta^{-1}$. Therefore,
in dimensionless units the boundary layer thickness is
\begin{equation}
\label{capalimit}
\delta(\theta) = \frac{1}{N_g\sin\theta}.
\end{equation}
This thickness increases when the particle approaches the line
$\theta=0$. For small angles the surface of contact between the
particles becomes vertical, and the rolling picture is no longer
applicable. Instead, we can imagine that particles diffuse away from
the fixed sphere at a certain angle, $\theta_0$. This separation is
expected to occur when the $x$ coordinate of the center of the
falling particle is beyond the region excluded by the fixed one.
Given eq.(\ref{capalimit}), this implies
$\left[1+\delta(\theta_0)\right]\cos\theta_0 \simeq 1$. Solving this
equation for small angles, we obtain $\theta_0 \simeq
\left(2/N_g\right)^{1/3}$. From this point the incident particle
falls a distance $z_0 \simeq \left(2/N_g\right)^{1/3}$; the
dimensionless time needed to reach the $z=0$ line is $t_0 \simeq
z_0/N_g = 2^{1/3}N_g^{-4/3}$. Due to the horizontal Brownian motion
performed during the particles' falling, one expects a horizontal
deviation of order $\delta x \simeq
\sqrt{2t_0} \simeq (2/N_g)^{2/3}$. This simple argument gives an
exponent and order of magnitude that agrees with the more exact
singular perturbation result (\ref{distanciab}).

\subsection{Brownian Dynamics Simulations}

Computer simulation allows us to study the adsorption probability for
the whole range of $N_g$ values and to verify the validity of the
results obtained in the previous subsections. The algorithm developed
here will be also used in the following section to study the
structure of jammed lines.

We consider, as described in section \ref{sec:description}, a
suspension of Brownian particles that is dilute enough to assume
sequential adsorption. One by one, the center-of-mass of a new
particles starts its motion at a fixed vertical position $z_0$ and at
a horizontal position $x_0$ chosen at random from a uniform
distribution. The value of $z_0$ does not affect the final adsorption
probability as far as in the initial position there is no possibility
of interaction with the adsorbed particles, that is, $z_0 > 1$. In
our simulations we take $z_0 = 1$ to minimize the simulation time
required.

The particle motion is discretized in time in the following way. At
every time step $\Delta t$ the particle develops two motions
performed sequentially in the simulations. It travels a vertical
distance $\Delta z_{det}=-N_g\Delta t$ (in the dimensionless units
introduced in section \ref{sec:description}) as corresponds to the
sedimentation under gravity and a stochastic displacement $\Delta
\vec r_{rd}$ as corresponds to the Brownian motion. Therefore,
\begin{equation}
\label{desplacament}
\Delta \vec r = -N_g\Delta t\hat z + \Delta\vec r_{rd}.
\end{equation}
As is well known, the Brownian displacement in two dimensions follows
a normal distribution law with zero mean and variance
\begin{equation}
\label{segonmoment}<\Delta \vec r_{rd}^{\,2}>=4\Delta t.
\end{equation}
After these two motions are generated, the simulation time is
increased by $\Delta t$ and the algorithm is repeated until the
particle reaches the adsorbing line $z=0$. The time step $\Delta t$
is chosen in such a way that ensures a typical displacement of the
particle to be of magnitude $\epsilon$ which is small compared to the
particle diameter, $\epsilon \ll 1$. The typical Brownian
displacement is of order $\sqrt{\Delta t}$, so $\Delta t \le
\epsilon^2$. On the other hand, the deterministic displacement is
$-N_g\Delta t$, so $\Delta t \le \epsilon/N_g$. Therefore, we impose
\begin{equation}
\label{pastemps}
\Delta t = {\rm min}\left[\epsilon^2,\epsilon/N_g\right].
\end{equation}
The value of $\epsilon$ is restricted by computer time limitations.
All the simulations reported in this paper were obtained with
$\epsilon=0.05$.

To complete the simulation algorithm, we need a collision rule to use
in the case that an incoming particle touches the pre-adsorbed one.
We consider two collision rules, depending on whether at the
collision instant the particle is performing a deterministic or a
random displacement. If the collision occurs during the deterministic
displacement, we consider that the incoming particle rolls over the
preadsorbed one. The collision rule for the Brownian motion is a
perfect reflection of the particle trajectory, as in an elastic
collision with an infinitely massive particle. This rule is an
approximation, valid if the Brownian displacement is sufficiently
small in relation to the radius of the boundary.

When the center of the particle reaches the $z=0$ line, we consider
that the particle adsorbs at the contact point, which is recorded.
By iteration of this algorithm one obtains a histogram of adsorption
frequencies as a function of the distance to the fixed particle.

For small values of $N_g$ the results obtained by Brownian dynamics
simulations agree with the analytical results obtained from the
superposition approximation eq.(\ref{superposicio}), and the
multipolar expansion eq.(\ref{multipolar}), see figure \ref{part1}.
For large values of $N_g$ the simulated distribution of adsorbed
particles approaches the scaling (\ref{scaling}) predicted by the
asymptotic approximation, see figure \ref{bl}.

To obtain a simple analytic approximation in the intermediate to high
$N_g$ regime, we fit the simulation results for $\rho(x)$ with an
exponential distribution,
\begin{equation}
\label{exp}
\rho (x) \simeq 1 + be^{-b(|x|-1)},\;\;\;|x|>1.
\end{equation}
Although the asymptotic form of the adsorption probability obtained
in appendix \ref{ap:layer} is not exactly an exponential, this is a
simple fitting function. To determine the value of $b(N_g)$ from the
simulation results we impose the condition that the thickness of the
exponential has to be equal to the thickness of $\rho^{(1)}$ obtained
from the simulations. This is the condition of maximum likelihood for
an exponential distribution \cite{LAVENDA}, and it implies
\begin{equation}
\frac 1b = <|x|>_1-1.
\end{equation}
For large values of $N_g$ we expect, based on our asymptotic argument
(\ref{distanciab}), a dependence of the form $<|x|>_1=1+\alpha
N_g^{-2/3}$. To determine the constant $\alpha$ we computed the mean
distance of the adsorbed particles to the previously adsorbed one for
different $N_g$ numbers. The simulations were performed for
$N_g=60,80,100,140,200,400,600$ using $40,000$ particles in each
simulation and we obtained:
\begin{equation}
\label{fit}
b=\frac{N_g^{2/3}}\alpha,\;\;\;\alpha=\fitt\ldots,
\end{equation}
in agreement with the value $\pert$ obtained from the asymptotic
solution (\ref{distanciab}). In figure \ref{part1exp} we compare this
exponential fit, eqs. (\ref{exp}), (\ref{fit}) with the results of
the Brownian dynamic simulations for $N_g=60$.

Making use of the superposition hypothesis (\ref{superposicio}) and
the normalization condition (\ref{normaP}) the probability density of
adsorption in a gap can be expressed in the simple form:
\begin{equation}
\label{kernelexp}
p(h',h) = \frac{1+be^{-bh}+be^{-b(h'-h-1)}}{h'+1-2e^{-b(h'-1)}}.
\end{equation}
Although the exponential distribution does not reproduce all the
details of the simulated $\rho(x)$, it approaches its main
characteristics, and the corresponding kernel (\ref{kernelexp}) is
simple enough to allow a numeric evaluation of the integral equation
(\ref{int}).

\section{Structure of the adsorbed layer}
\label{sec:structure}

In this section we present results concerning the structure of the
adsorbed layer at jamming, characterized by the coverage
$\theta_\infty$ and the radial distribution function $g(r)$. In
subsection \ref{subsec:jamcoverage} we study the variation of the
jamming coverage with $N_g$ and in subsection \ref{subsec:g(r)} we
analyze the local structure of jammed surfaces described by $g(r)$.
The results obtained by numerical integration of eq.(\ref{int}) using
(\ref{kernelexp}) or (\ref{multipolar}) are compared with simulation
results.

The characteristics of the simulations are as follows. For each value
of $N_g$ we have filled $1000$ lines of length $L=100$ and we have
computed the mean number of particles adsorbed at saturation. The
statistical uncertainty on the coverage (error bars) is estimated by
a 95\% confidence interval. In principle, particle trajectories are
simulated as long as the surfaces are not entirely covered, but there
exist some special cases that demand more attention to avoid
prohibitive computer time. Consider a particle diffusing in the
region between two pre-adsorbed ones, being the interval of the line
between these pre-adsorbed particles of length less than $1$. This
particle only can adsorb if it obtains sufficient thermal energy to
overcome the gravitational force and the geometrical barrier
delimited by the pre-adsorbed particles. The time needed to escape
from these ``geometrical barriers'' grows rapidly with $N_g$. For
$N_g\sim 1$ the particle can return to the bulk after a small time
and try to adsorb elsewhere. However, for sufficiently high $N_g$
this time becomes very large (infinite in the case of BD). In the
practical situation, these particles play no role because other
particles, coming from the bulk, will adsorb in the free gaps. There
are different ways in which this effect can be taken into
account\cite{RAED}. In our simulations we adopt the following
criteria: we choose a maximum residence time $\tau$ and, if a given
particle remains time $t>\tau$ in a trap, it is eliminated from
our simulations and a new particle starts the motion following the
previous rules. We take a different characteristic residence time
$\tau$ for each $N_g$. For small $N_g$ only a few particles are
eliminated, but for $N_g \gg 1$ usually all particles that fall into
intervals of length $h<1$ are eliminated. For $R^*\leq1$ we take
$\tau=15$, $\tau=8$ for $1<R^*\leq2$ and $\tau=1$ for $R^*>2$. We
have checked that our results are not sensible to increments of
$\tau$. Finally, the last stage is simulated in the following way.
When only gaps with length $h'<3$ survive, the initial position $x_0$
is chosen at random between the centers of the particles limiting
these gaps, at $z=1$ as for the other particles. When only targets of
length $h' < 2$ remain (at most one particle can be adsorbed in each)
we fill these by using a RSA algorithm, i.e. by randomly placing a
particle in it without any transport process. Only a few particles
are deposited following these two last rules and therefore the
structure (characterized by $g(r)$) is not significantly altered.

\subsection{Coverage at the jamming limit}
\label{subsec:jamcoverage}

For adsorption on an infinite line, the saturation coverage can only
be a function of the unique dimensionless parameter present in the
transport equation, namely $N_g$. It is more convenient to represent
the variation of $\theta_\infty$ as a function of the dimensionless
radius, $R^*$, defined in eq. (\ref{R*}), which gives a more compact
scale than $N_g$.

In figure \ref{recPe} we show the variation of $\theta_\infty$ with
$R^*$ obtained by simulation (crosses) and by solving the integral
equation (\ref{int}) following the numerical method described in
\cite{DRSAB}. The solid line in the region $R^*<1.5$ ($N_g\le 10$)
has been obtained with the kernel given by substitution of the
multipolar expression (\ref{multipolar}) truncated to five terms in
eq. (\ref{superposicio}). We also solve, for all values of $R^*$, the
integral equation (\ref{int}) with the kernel obtained using the
exponential approximation, eq. (\ref{kernelexp}); the result is
represented by the solid line for $R^*>1.5$ and by the short dashed
line for $R^*<1.5$.

Three main regions can be recognized in the graph:

\begin{itemize}
\begin{enumerate}
\item $R^*\leq 1$: In this region the Brownian motion dominates. The
coverage varies slowly with $R^*$ and its value remains close the
DRSA value ($\theta_\infty \simeq 0.7506$). For example, for the
$R^{*}=1$ case we obtain $\theta_\infty \simeq 0.7539$ by means of
simulations and the numerical solution of the integral equation with
the multipolar kernel yields $\theta_\infty\simeq 0.7528$.
Simulation results agree with the curve obtained using the multipolar
solution of transport equation. The curve obtained using the
exponential kernel (\ref{kernelexp}) gives smaller values for
$\theta_\infty $ and in the limit of vanishing gravity it approaches
the RSA coverage $\theta_\infty\simeq \rsa$ instead of the DRSA
value.

\item $1\leq R^{*}\leq 2.5$: $\theta_\infty $ rises quickly with
$R^{*}$. In this region the numerical solutions of the integral
equation with the multipolar expression and with the exponential
approximation (\ref{kernelexp}) are close. However, they show a
slight discrepancy with the simulation results that are
systematically greater.

\item $R^{*}\ge 2.5$ $\theta_\infty $ approaches slowly to the BD
limit ($\theta_\infty^{BD}=\bd$), for example, if we take $R^{*}=10$
($N_g=10^4$) simulations give $\theta_\infty =0.807...$. We obtain a
good agreement between the simulations and the solution of
(\ref{int}) using the exponential kernel (\ref{kernelexp}). A simple
expression describes the asymptotic approach to the strong gravity
limit, $N_g\to \infty$:
\begin{equation}
\label{AsimBD}
\theta_{\infty}(N_g\gg1)\simeq\theta_{\infty}^{BD}-\asint N_g^{-2/3}.
\end{equation}
This expression is shown in fig. \ref{recPe} by dashed lines, and can
be obtained considering a simplified model. For large values of
$N_g$, the explicit form of $\rho^{(1)}$ seems not to be of great
importance provided that it gives a value for the mean distance
between the adsorbed particles and the pre-adsorbed one in agreement
with eq.(\ref{distanciab}). We can use instead of the exact density,
an approximation in which $\rho^{(1)}$ is described by a normalized
step function of thickness $2<|x|>_1 = 2AN_g^{-2/3}$ in agreement
with (\ref{distanciab}). This model is exactly solvable \cite{Futur},
giving (\ref{AsimBD}) in the limit $N_g\to\infty$.
\end{enumerate}
\end{itemize}

We have performed some control simulations to study the origin of the
slight discrepancy between the solid line and simulations in region 2
on the graph. This effect is due to the fact that for this range of
$N_g$ the probability density of adsorption on a gap not only depends
on the gap's limiting particles but can also be influenced by third
neighbors. This can be clearly shown by performing simulations with
three pre-adsorbed spheres and comparing the resulting adsorption
density probability with the result of the superposition
approximation (\ref{superposicio}). In figure \ref{3part} we show the
adsorption probability density in a situation with three pre-adsorbed
spheres on a line of length $L=20$ and $R^{*}=1$. Between spheres one
and two there is a gap of length $h'=2.5$ and between spheres two and
three there is a small gap of $h'=0.05$ that does not allow particle
adsorption therein. Simulation results show an asymmetric adsorption
probability, which clearly displays the effect of the third particle.
The solution corresponding to the superposition hypothesis
(\ref{superposicio}), which assumes that the adsorption probability
in one gap is given by the product of the one particle solutions
corresponding to the particles at the ends of the interval, shown in
the continuous line, is symmetric. We also note that the simulation
results fit well with a distribution constructed as a product of the
three one-particle solutions (dashed line). For larger $R^{*}$
values, the ansatz of independence of gaps holds, and the theoretical
expression shows no significant deviation from their simulation
counterpart.

\subsection{Radial distribution function at the jamming limit}
\label{subsec:g(r)}

The radial distribution function, $g(r)$, reflects the correlation
existing between the adsorbed particles. The definition adopted here
is the usual one in Statistical Mechanics (see for instance
\cite{HANSEN}). In this subsection, we present histograms of $g(r)$
in the jamming limit, obtained from the Brownian dynamic simulations
used in the previous subsection to determine $\theta_\infty $.

In fig.\ref{satgr1} and \ref{satgr2} we show $g(r)$ for various
values of $R^*$. At the jamming limit, $g(r)$ presents a logarithmic
singularity at $r=1$, like the RSA model. The simulation results,
however, are histograms that have a finite value at contact
representing an average of $g(r)$ over a finite distance interval.
This contact value is nearly constant for $R^*\le1$ whereas it grows
with $R^*$ for $R^*>1$.

For $R^{*}\le 1$ our $g(r)$ are indistinguishable from the results
obtained for DRSA ($R^*=0$). The $g(r)$ for $R^{*}=1$ is shown in
fig. \ref{satgr1}, note the disappearance of correlations between
particles after a few diameters. We recall that the variation of
$\theta_\infty $ with $R^{*}$ in this region is less than $1\%$, as
we pointed out in the previous section. Therefore, we can conclude
that for $R^{*}\le 1$ the structure of the adsorbed layer is almost
independent of gravity.

In fig. \ref{satgr2} we show the radial distribution function for
$N_g=30 \quad (R^*=1.968)$ and for $N_g=240 \quad (R^*=3.310)$. A
comparison between figures \ref{satgr1} and \ref{satgr2} shows that
peaks in $g(r)$ increase and are steeper as $R^*$ increases,
revealing the tendency of large particles to pack closer together
than smaller ones. This is because diffusion is a small effect for
large particles, and there is an increased probability of adsorption
close to already adsorbed particles. We can observe that the smaller
the effect of gravity (as measured by $R^*$), the poorer the
structure in the radial distribution function. For $1 \le R^*\le 2.5$
not only the coverage changes rapidly with $R^*$ (as noted in
subsection \ref{subsec:jamcoverage}) but $g(r)$ does also. Note for
example that, for $R^*=1.968$, two secondary peaks are visible
whereas only one is visible for $R^*=1$. For $R^*>2.5$ the obtained
$g(r)$ are very close to the correlation function of the ballistic
deposition model, although the delta function singularities present
in the BD model at $r=1,2,\ldots$ are smoothed due to the effect of
diffusion. Our model differs from other generalized models
\cite{BALgen} in which the delta funciton singularities appear only
for $R^*=\infty$.

At the jamming limit $g(r)$ cannot be obtained completely using the
formalism presented in section \ref{sec:description}, but it can be
obtained for $1 \leq r \leq 2$ noting that $n(h)$ is equal to
$g(h+1)$ for $0\leq h\leq 1$ except by the normalization. By the
definition of these functions one has
\begin{equation}
\label{relng}
n(h) = \theta_\infty^2g(h+1),\;\;\;h<1.
\end{equation}
The numeric solution of eq.(\ref{int}) performed in the previous
subsection to find $\theta_\infty$ also gives the gap density $n(h)$.
The result agrees with the radial distribution function obtained by
simulation according to eq.(\ref{relng}), as shown in fig.
\ref{satgr1} for $R^*=1$ and in fig. \ref{satgr2} for
$R^*=1.968,\;3.310$.

\section{Conclusions}

A 1+1 dimensional adsorption model has been analyzed to investigate
the influence of transport mechanisms (diffusion and sedimentation)
on the structure of irreversibly adsorbed monolayers of particles.
We have studied a simplified model in which hard spheres suspended in
a two-dimensional fluid adsorb onto a line. The results may be
relevant to understand the more realistic case of adsorption on a
surface, if the suggested scaling behavior between the 1+1 and 2+1
dimensional cases holds \cite{Lleiescala}. An argument in favor of
this hypothesis for large gravity is presented below.

By means of Brownian dynamic simulations we have obtained the
saturation coverage and the radial distribution function $g(r)$ for
different values of the gravity number $N_g$. These results have been
compared with those obtained from an approximate analytic formalism:
first we analyze the probability density of adsorption onto a gap
with the help of the transport equation and Brownian dynamic
simulations; once we know this probability distribution, the
saturation coverage $\theta_\infty (R^{*})$ and the density of gaps
at the jamming limit can be obtained numerically by using an integral
equation developed in a previous paper \cite{DRSAB}.

We have introduced a superposition hypothesis, assuming that the
adsorption probability at a given point depends only on the limiting
particles of the gap in which the particle adsorbs. As showed by
simulations, this hypothesis fails when the gap is small and $1\le
R^*\le2$, but this affects $\theta_\infty$ only slightly.

The probability density of adsorption in the presence of one particle
was obtained by solving the transport equation as a multipolar
expansion. This expansion is useful for low to intermediate $N_g$
values, but not for high $N_g$ numbers. In the large $N_g$ regime we
have shown, using singular perturbation techniques, that the
probability density of adsorption $\rho^{(1)}(x)$ has the scaling
form (\ref{scaling}). The simulations show that this scaling is
approximately satisfied even for moderate values of $N_g$. Instead of
the exact scaling function, we have useda simple exponential function
for the numerical computations that fits the simulation results. The
corresponding numeric solution of the integral equation (\ref{int})
agree well with the simulations even for $R^*>1$.

The properties of the adsorbed layers at jamming are as follows. For
$R^*\le1$, $\theta_\infty(R^{*})$ varies less than $1\%$ and $g(r)$
is close to the DRSA ($R^{*}=0$) form. Therefore, the effect of
gravity is nearly negligible in this regime. For $R^*\ge1$,
$\theta_\infty(R^*)$ grows quickly with $N_g$ and $g(r)$ displays
peaks that emerge more significantly. The peaks in $g(r)$ increase
and are steeper when $R^{*}$ grows reflecting the tendency of large
particles to pack closer than smaller ones. For $R^{*}\ge 2.5$ the
saturation coverage slowly approaches the BD value following the
asymptotic expression (\ref{AsimBD}). In this regime $g(r)$ is very
close to its BD counterpart, although in our model $g(r)$ takes
finite values at $r=1,2,\ldots$ due to the effect of diffusion
instead of the delta function singularities of the BD case.

It can be easily verified that the same perturbation techniques used
in appendix \ref{ap:layer} can be applied to the $2+1$ dimensional
case, leading to the scaling given in eq. (\ref{scaling}) with the
same function $\phi^{(1)}$. This fact provides a physical basis, at
least for large gravity, for the scaling of $\theta_\infty (R^{*})$
in a common curve for the $2+1$ and $1+1$ cases. In the asymptotic
limit the relevant property of the adsorption probability is its
thickness, and one can replace $\rho^{(1)}$ with a step function with
the same thickness, $\Delta=k N_g^{-2/3}$. We expect that the jamming
coverage approaches the ballistic limit linearly in $\Delta$, and
therefore
\begin{equation}
\frac{\theta_{\infty}(N_g\gg1)}{\theta_{\infty}^{BD}}\simeq
1-\alpha_d N_g^{-2/3},
\end{equation}
where $\alpha_d$ is a constant that depends on the dimension of the
system. Thus, with an adequate change of scale the curves of
$\theta_\infty(N_g)/\theta_{\infty}^{BD}$ in 1+1 and 2+1 dimensions
are coincident, at least for large values of $N_g$. This scaling has
been observed in simulations and seems to apply for all values of
$N_g$ \cite{Lleiescala}.

In a more realistic extension of our model, hydrodynamic interactions
must be considered in the transport equation, but we expect that
their effect is small. In the DRSA case, hydrodynamic interactions
\cite{HYDRO} make the adsorption probability nearly uniform due to
the enhanced mobility parallel to the surface and therefore the
coverage is expected to be close to the RSA case. In the case of BD
it has been shown \cite{BDHYDRO} that while the jamming coverage does
not change significantly, the local structure is strongly affected by
the hydrodynamic interactions.

\acknowledgements
Useful comments by Ignacio Pagonabarraga are acknowledged. J.F. is
supported by a doctoral scholarship from the Programa de formaci\'o
d'investigadors of the Generalitat de Catalunya under grant
FI/96-2.683. We also acknowledge financial support from the
Direcci\'on General de Investigaci\'on of the Spanish Ministry of
Education and Science (grant PB94-0718) and the European Union under
grant ERBCHRXCT 920007.

\appendix

\section{Multipolar Solution of Transport Equation}
\label{ap:multi}

In this appendix we obtain $\psi^{(1)}(\vec r)$ using the Green
function method. The equation and boundary conditions for $\psi
^{(1)}(\vec r)$ can be obtained by substitution of
(\ref{dosparts}),(\ref{fi0}) in
(\ref{Smolu2}),(\ref{contorn1}),(\ref{contorn2}). If we write
$\psi^{(1)}(\vec r)=f(\vec r)e^{-N_gz/2}$, the equation for $f(r)$ is
\begin{equation}
\label{eqf}
\nabla^2 f(\vec{r})= \frac{N_g^2}{4} f(\vec{r}),
\end{equation}
with boundary conditions:
\begin{eqnarray}
\label{contornf}
f(x,z=0) &=& 0 \\
\left. \frac{\partial f(\vec{r})}{\partial r}\right|_{r=1} +
\frac{N_g}{2}\sin\theta f(r=1,\theta) &=&
-\sin\theta e^{\frac{N_g}2\sin\theta}.
\end{eqnarray}
The Green function corresponding to (\ref{eqf}) and satisfying the
boundary condition (\ref{contornf}), $G(\vec r,\vec r\, ')$, is
defined by
\begin{eqnarray}
\label{eqG}
\nabla^2G(\vec{r},\vec{r}\,') - \frac{N_g^2}{4}G(\vec{r},\vec{r}\,')
&=& \delta (\vec{r}-\vec{r}\, ') \\
G(x,z=0,\vec{r}\,') &=& 0.
\end{eqnarray}
Using the method of images, it is easy to obtain the solution to this
equation, that can be expressed using Bessel functions of the third
kind:
\begin{eqnarray}
\label{G1}
&&G(\vec{r},\vec{r}\,') = \frac{-1}{2\pi}\left[ K_0 \left(
\frac{N_g}{2} \vert \vec{r}-\vec{r}\,' \vert \right)- K_0 \left(
\frac{N_g}{2} \vert \vec{r}-\vec{r}^{\, *} \vert \right) \right] \\
&&\vec{r}=(x,z) \quad \vec{r}\, '=(x',z') \quad \vec{r}^{\, *} =
(x',-z').\nonumber
\end{eqnarray}

We can obtain an integral equation for $f(\vec r)$ making use of the
properties of the Green function. If we multiply eq. (\ref{eqf}) by
$G(\vec r,\vec r\, ')$ and eq. (\ref{eqG}) by $-f(\vec r)$ and add
the results, we obtain:
\begin{equation}
\label{Gf}
G(\vec{r},\vec{r}\, ') \nabla^2 f(\vec{r}) - f(\vec{r}) \nabla^2
G(\vec{r},\vec{r}\, ') = -f(\vec{r}) \delta (\vec{r}-\vec{r}\, ').
\end{equation}
Now, integrating (\ref{Gf}) over the whole volume $V$, limited by the
adsorbing line at $z=0$, the excluded volume of the pre-adsorbed
particle and a closing surface $S$, and then using the Stokes
theorem, we obtain:
\begin{eqnarray}
\label{Gf2}
f(\vec{r}\, ') &=& -\int_VdV \left[ G(\vec{r},\vec{r}\, ') \nabla^2
f(\vec{r}) - f(\vec{r}) \nabla^2 G(\vec{r},\vec{r}\, ') \right]
\nonumber\\
&=& -\int d\vec{S} \left[ G(\vec{r},\vec{r}\, ')
\vec{\nabla} f(\vec{r}) \right] +\int d\vec{S}
\left[ f(\vec{r})\vec{\nabla}G(\vec{r},\vec{r}\, ') \right].
\end{eqnarray}
If the surface S goes to infinity and we use the boundary conditions
for $f$ and $G(\vec r,\vec r\, ')$, the only contribution to the
surface integrals comes from the excluded area of the preadsorbed
particle. Therefore, using $dS=Rd\theta =d\theta $ we obtain:
\begin{equation}
\label{Ap1}
f(\vec{r}\, ') = \int_0^\pi d\theta \left[ G(\vec{r},\vec{r}\, ')
\frac{\partial f}{\partial r}-f(\vec{r})\frac{ \partial G}
{\partial r} \right]_{r=1},\;\;\; r'>1.
\end{equation}

To take profit of this expression, we use the Fourier expansion of
the Green function (\ref{G1}) that can be obtained using a well-known
theorem for the Bessel functions \cite{ABRAMOWITZ}:
\begin{equation}
\label{ThB}
K_0 \left(\beta \sqrt{r_1^2+r_2^2-2r_1r_2\cos\phi}\right) =
\sum_{m=-\infty}^{\infty} K_m(\beta r_2) I_m(\beta r_1)\cos m\phi,
\quad r_2 \ge r_1.
\end{equation}
Use of (\ref{ThB}) in (\ref{G1}) yields: 
\begin{eqnarray}
\label{G}
&&G(\vec{r},\vec{r}\, ') = \frac{-2}{\pi} \sum_{n=1}^\infty K_n
\left( \frac{N_g}{2} r' \right) I_n \left( \frac{N_g}{2}r \right)
\sin(n\theta)\sin(n\theta'), \quad r \le r' \\
&&x=r\cos\theta \quad z = r\sin\theta \quad x' = r'\cos\theta' \quad z'
= r'\sin\theta'.\nonumber
\end{eqnarray}
Substitution of this expression in eq. (\ref{Ap1}) yields
\begin{eqnarray}
\label{Ap2}
f(\vec{r}\, ') &=& -\frac{2}{\pi} \sum_{n=1}^\infty
K_n(\frac{N_g}{2}r')I_n(\frac{N_g}{2})\sin(n\theta')
\int_0^\pi d\theta\sin(n\theta)\frac{\partial f}{\partial r}_{r=1}
\nonumber\\
&+& \frac{2}{\pi} \sum_{n=1}^\infty K_n(\frac{N_g}{2}r')\frac{N_g}{4}
\left[I_{n+1}(\frac{N_g}{2}) + I_{n-1}(\frac{N_g}{2})\right]
\sin(n\theta')\int_0^\pi d\theta f(r=1, \theta)\sin(n\theta).
\end{eqnarray}
Now, we note that our problem is symmetric under the change $x\to
\pi-x$, and, therefore, only the integrals with odd $n$ contribute.
In addition, if we apply the boundary condition of null radial flux
at the excluded area defined by the preadsorbed sphere
(\ref{contornf}) and define $\varphi (\theta )=f(r=1,\theta )$ we
obtain
\begin{eqnarray}
\label{Ap3}
f(\vec{r}\, ') &=& \frac{2}{\pi} \sum_{n=1}^\infty
K_{2n-1}(\frac{N_g}{2}r') \sin(2n-1)\theta '\int_0^\pi d\theta
\sin(2n-1)\theta \left[\sin\theta e^{\frac{N_g}{2}\sin\theta}
I_{2n-1}(\frac{N_g}{2}) + \right. \nonumber\\
&& \left. \left(\frac{N_g}{2} I_{2n-1}(\frac{N_g}{2})\sin\theta +
\frac{N_g}{4} (I_{2n-2}(\frac{N_g}{2})+I_{2n}(\frac{N_g}{2})) \right)
\varphi(\theta)\right].
\end{eqnarray}
If we define the Fourier-sine transforms:
\begin{eqnarray}
\label{Ap4}
g_{2n-1}(N_g) &=& \int_0^\pi d\theta \sin(2n-1)\theta \sin\theta
e^{\frac{N_g}{2}\sin\theta}, \\
\Gamma_{2n-1}(N_g) &=& \int_0^\pi d\theta \sin(2n-1)\theta \sin\theta
\varphi(\theta), \\
\label{Ap41}
\varphi_{2n-1}(N_g) &=& \int_0^\pi d\theta\sin(2n-1)\theta
\varphi(\theta),
\end{eqnarray}
finally one obtains for $f(r,\theta )$ (we omit ' for simplicity): 
\begin{equation}
\label{Ap5}
f(r,\theta)=\frac{2}{\pi}\sum_{n=1}^{\infty}\varphi_{2n-1} \frac{
K_{2n-1}(\frac{N_g}{2}r)}{K_{2n-1}(\frac{N_g}{2})}\sin(2n-1)\theta.
\end{equation}
By substitution of $r=1$ in (\ref{Ap3}) and making use of definitions
(\ref{Ap4}-\ref{Ap41}) we obtain the following equations for the
coefficients $\varphi_{2n-1}$:
\begin{eqnarray}
\label{Ap6}
\varphi_{2n-1} &=& K_{2n-1}(\frac{N_g}{2}) I_{2n-1}(\frac{N_g}{2})
\frac{g_{2n-1}+\frac{N_g}{2}\Gamma_{2n-1}}{1-\frac{N_g}{4}K_{2n-1}
(\frac{N_g}{2})(I_{2n-2}(\frac{N_g}{2})+I_{2n}(\frac{N_g}{2}))}, \\
\Gamma_{2n-1} &=& \frac{2}{\pi}\sum_{m=1}^{\infty}
\frac{-4(2m-1)(2n-1)}{(2n+2m-3)(2n+2m-1)(2n-2m+1)(2n-2m-1)}
\varphi_{2m-1}. \label{Ap61}
\end{eqnarray}
The probability density for the adsorption of a particle can be
obtained in a form of a multipolar expansion by using eq.(\ref{Ap5})
with the definition of $f(r)$, $\psi^{(1)}(\vec r)=f(\vec
r)e^{-N_gz/2}$ in definitions (\ref{dosparts}), (\ref{ro3}):
\begin{equation}
\label{Ap7}
\rho(x)=1 + \frac{2}{\pi \vert x \vert} \sum_{n=1}^{\infty}
(2n-1)\varphi_{2n-1} \frac{K_{2n-1}(\frac{N_g}{2}
\vert x \vert)}{K_{2n-1}(\frac{N_g}{2})},\;\;\;|x|>1.
\end{equation}
For small values of $N_g$, one has $g_{2n-1} =
\frac{\pi}{2}\delta_{n1} + {\cal O}(N_g)$, and all the terms in
eq.(\ref{Ap7}) except the first vanish to order ${\cal O}(N_g)$,
giving $\varphi_{2n-1} = \frac{\pi}{2}\delta_{n1} + {\cal O}(N_g)$.

\section{Boundary Layer Solution of the Transport Equation}
\label{ap:layer}

Our aim in this appendix is to obtain an approximate solution of the
transport equation (\ref{Smolu2}) valid for large values of $N_g$.
When $N_g$ is very large, the laplacian term in the transport
equation becomes negligible in the entire domain except near the
boundaries, where the value of the derivatives can be large. We have
then a singular perturbation problem\cite{KEVORKIAN}. An approximate
solution in this limit can be reached by matching different
approximations in the bulk and in the boundary layers that can appear
near the boundaries.

The first approximation to the {\em outer solution} can be obtained
neglecting the laplacian term in the transport equation. The
resulting equation is simply $\partial\psi/\partial{z}=0$, and the
solution satisfying the boundary condition at infinity eq.
(\ref{contorn1}) is a constant,
\begin{equation}
\label{outer}
\psi_{out} = 1/N_g.
\end{equation}
Note that this is an exact solution of the transport equation, and
therefore the expansion of the outer solution ends here.

This outer solution does not match the boundary conditions at $z=0$
and $r=1$, and therefore some boundary layers have to be introduced.

{\bf (I).} A first boundary layer, appears near the line $z=0$, where
the value of $\psi$ has to change from the outer value, $1/N_g$, to
the value imposed by the boundary condition, $0$. Derivatives with
respect to $z$ become large, and the term
$\partial^2\psi/\partial{z^2}$ in the laplacian is no longer
negligible. The boundary layer thickness, $\epsilon$, can be
estimated from the condition that the derivatives with respect to $z$
are the dominant terms in the transport equation and have the same
order. This implies $\epsilon = {\cal O}(N_g^{-1})$. Therefore, the
first approximation to the {\em inner solution} in this region is the
solution of
\begin{equation}
\label{layerI}
\frac{\partial^2\psi^I_{in}}{\partial{z^2}} +
N_g\frac{\partial\psi^I_{in}}{\partial{z}}=0.
\end{equation}
The solution satisfying the boundary condition at $z=0$ and matching
the outer solution for $z\gg N_g^{-1}$ is
\begin{equation}
\psi^I_{in} =\frac 1{N_g}\left(1-e^{-N_gz}\right).
\end{equation}
This is also an exact solution of the transport equation, so the
expansion of the inner solution in this boundary layer ends here. In
the absence of any adsorbed particle, no additional boundary layers
are present and, in fact, we have reproduced the exact solution for
an empty line given by eq. (\ref{fi0}).

{\bf (II).} In the presence of one particle in $r=0$, a new boundary
layer appears near the boundary of that particle, $r=1$. The function
$\psi$ has to change in a small boundary of thickness $\delta$ from
the outer form (\ref{outer}), in which there is a radial probability
flux, to a form in which that flux vanishes at $r=1$. The order of
magnitude of the different terms of the transport equation can be
easily evaluated by writing it in polar coordinates, and considering
that each radial derivative introduces a factor ${\cal O}
(\delta^{-1})$:
\begin{eqnarray}
\label{Ogran}
&\frac{\gran \partial^2\psi}{{\gran \partial{r}}^2}& + 
\frac{\gran 1}{\gran r}\frac{\gran \partial\psi}{\gran \partial{r}} + 
\frac{\gran 1}{{\gran r}^2}\frac{\gran \partial^2\psi}{\gran\partial\theta^2}
+ N_g\sin\theta\frac{\gran \partial\psi}{\gran \partial{r}} + 
N_g\frac{\gran \cos\theta}{\gran r}\frac{\gran \partial\psi}{\gran
\partial\theta}
=  0. \cr
&{\cal O}(\delta^{-2})& \quad {\cal O}(\delta^{-1}) \quad  {\cal O}(1) 
\qquad {\cal O}(N_g\delta^{-1}) \qquad {\cal O}(N_g) 
\end{eqnarray}
The first and fourth terms of this equation are the largest when
$N_g\to\infty$, and the appropriate {\em distinguished limit} is
obtained when they have the same order of magnitude, that is, when
$\delta={\cal O}(N_g^{-1})$. To obtain the solution in this boundary
layer, we define the scaled inner variable $\eta\equiv N_g(r-1)$, and
we expand the function $\psi^{II}_{in}$ in successive approximations,
$\psi^{II}_{in}=\psi^{(1)} + \psi^{(2)}+\cdots$. The first
approximation to the inner solution is then, retaining the terms of
order ${\cal O}(N_g^2)$,
\begin{equation}
\label{layerII1}
\frac{\partial^2\psi^{(1)}}{\partial{\eta}^2} + 
\sin\theta\frac{\partial\psi^{(1)}}{\partial{\eta}} = 0.
\end{equation}
The boundary conditions for $\psi^{(1)}$ cause the normal flux to
vanish at $\eta=0$ and the solution has to match the outer solution
for $\eta\to\infty$. The first integral of eq. (\ref{layerII1}) is
immediate,
\begin{equation}
\frac{\partial\psi^{(1)}}{\partial{\eta}} + 
\sin\theta\psi^{(1)} = C(\theta).
\end{equation}
The term on the right-hand side is an integration constant that, in
principle, can be an arbitrary function of $\theta$. However, the
left hand side is proportional to the radial flux and, by the
boundary condition at $\eta=0$, the integration constant must vanish.
We have, then, a homogeneous first-order equation for $\psi^{(1)}$,
whose general solution is
\begin{equation}
\label{solII1}
\psi^{(1)} = A(\theta)e^{-\eta\sin\theta}.
\end{equation}
It is not possible to obtain the ``integration constant'' $A(\theta)$
by direct matching with the outer solution, because this first
approximation to the inner solution vanishes when $\eta\to\infty$. As
we will see, the reason for this is that the first approximation
$\psi^{(1)}$ is a quantity of order ${\cal O}(1)$, whereas the outer
solution is ${\cal O}(N_g^{-1})$. Matching is possible only for the
second-order approximation, $\psi^{(2)}$, which is ${\cal O}
(N_g^{-1})$. Retaining terms of order ${\cal O}(N_g)$ in the
transport equation we obtain the equation satisfied by $\psi^{(2)}$,
\begin{eqnarray}
\label{layerII2}
\frac{\partial^2\psi^{(2)}}{\partial{\eta}^2} & + & 
\sin\theta\frac{\partial\psi^{(2)}}{\partial{\eta}} = 
-N_g^{-1}\left(\frac{\partial\psi^{(1)}}{\partial{\eta}} + 
\cos\theta\frac{\partial\psi^{(1)}}{\partial{\theta}}\right) \nonumber\\
& = & -N_g^{-1}\cos\theta{e}^{-\eta\sin\theta}
\left[A'(\theta) - A(\theta)(\eta\cos\theta + \tan\theta)\right].
\end{eqnarray}
The first integral satisfying the boundary condition at $\eta=0$ is
\begin{equation}
\frac{\partial\psi^{(2)}}{\partial{\eta}} + 
\sin\theta\psi^{(2)} =
-N_g^{-1}\cos\theta\int_0^{\eta} d\eta'{e}^{-\eta'\sin\theta}
\left[A'(\theta) - A(\theta)(\eta'\cos\theta + \tan\theta)\right].
\end{equation}
It is not necessary to go beyond this calculation to obtain the
matching condition with the outer solution. It suffices to recognize
that the left-hand side of the last equation is $-N_g^{-1}$ times the
radial probability flux, and this quantity has to match when
$\eta\to\infty$ with the outer value for this flux, which from
(\ref{outer}) is simply $-\sin\theta$. We obtain the condition:
\begin{equation}
-\sin\theta = \cos\theta\left\{
\frac{A'(\theta)-A(\theta)\tan\theta}{\sin\theta} -
\frac{A(\theta)\cos\theta}{\sin^2\theta}\right\}.
\end{equation}
This is a differential equation for the unknown function $A(\theta)$,
whose general solution is $A(\theta)=\sin\theta + B\tan\theta$. The
arbitrary constant $B$ has to vanish in order to obtain a finite
probability density for $\theta=\pi/2$. This completely determines
the first approximation to $\psi^{II}_{in}$.

However, this approximate solution is not valid for all values of the
angle $\theta$. The reason is that the boundary layer thickness is
$\delta=(N_g\sin\theta)^{-1}$, as is evident from eq. (\ref{solII1}),
and it increases with decreasing angle. Therefore, the boundary layer
approximation breaks down when $\theta\to0$. It is easy to verify
that the order of magnitude of the last term on the left-hand side of
eq.(\ref{Ogran}), which has been neglected, becomes comparable to the
retained terms when $\theta={\cal O}(N_g^{-1/3})$. Therefore, the
range of validity of the solution obtained is
\begin{equation}
\label{solII2}
\psi^{II}_{in} = \sin\theta e^{-\eta\sin\theta} + {\cal O}(N_g^{-1}),
\;\;\;\; \theta\gg N_g^{-1/3}.
\end{equation}

{\bf (III).} For values of the angle $\theta={\cal O} (N_g^{-1/3})$,
the last term in eq. (\ref{Ogran}) has to be included in the boundary
layer equation together with the two terms retained in the previous
approximation. All these terms have the same order of magnitude when
the radial thickness of the boundary layer is $\delta={\cal O}
(N_g^{-2/3})$. These range of values for the angle and the radius
define a new boundary layer; we introduce new appropriately scaled
variables, $\xi=(r-1)N_g^{2/3}$, and $\tau=\theta N_g^{1/3}$. We
introduce also an asymptotic expansion of the probability density,
$\psi^{III}_{in} = \mu_1(N_g)\phi^{(1)} + \mu_2(N_g)\phi^{(2)} +
\cdots$, where the coefficients $\mu_i(N_g)$ are infinitesimals of
increasing order. The equation for $\phi^{(1)}$ is, retaining the
terms of larger order in $N_g$:
\begin{equation}
\label{layerIII}
\frac{\partial\phi^{(1)}}{\partial{\tau}} +
\frac{\partial^2\phi^{(1)}}{\partial{\xi}^2} +
\tau\frac{\partial\phi^{(1)}}{\partial{\xi}} = 0.
\end{equation}
We now have a parabolic equation that has the form of a time-reversed
diffusion equation with a drift proportional to the ``time'' $\tau$.
For $\tau\to\infty$, the solution has to match with the previous
boundary layer solution $\psi^{II}_{in}$, valid for large values of
the angle. It is easy to see that this implies
\begin{eqnarray}
\label{asymp}
\mu_1(N_g) &=& N_g^{-1/3}, \nonumber\\
\phi^{(1)}\to\tau e^{-\tau\xi}, && \;\;\;\tau\to+\infty.
\end{eqnarray}
That is, for large values of $\tau$ we recover a normalized
``Boltzmann distribution'', corresponding to neglecting the
$\tau$-derivative in eq. (\ref{layerIII}). This condition, together
with the boundary of condition of zero normal flux at $\xi=0$,
$\partial\phi^{(1)}/\partial\xi + \tau\phi^{(1)} = 0$, completely
specifies the function $\phi^{(1)}$. Furthermore, the condition of
zero flux at $\xi=0$, implies that the integral of $\phi^{(1)}$ with
respect to $\xi$ is a conserved quantity, and as a consequence of eq.
(\ref{asymp}) it has value $1$:
\begin{equation}
\label{normal}
\int_0^{\infty}\phi^{(1)}(\xi,\tau) d\xi = 1.
\end{equation}
We have not been able to analytically solve eq. (\ref{layerIII}).
However, it is easy to obtain a numeric solution satisfying the cited
boundary conditions. The solution is a positive normalized function,
vanishing exponentially for $\xi\to\infty$, and with a well-defined
limit when $\tau\to0$, $\phi^{(1)}(\xi,\tau=0)$ (see figure
\ref{bl}). An important quantity is its first moment,
$A=\int_0^{\infty}d\xi\xi\phi^{(1)}(\xi)=\pert$.

The adsorbing boundary condition at $z=0$, however, is still not
satisfied by this approximation. From the previous solution we have
$\lim_{\tau\to0}\psi = N_g^{-1/3}\phi^{(1)} (\xi,0)$ and this result,
obtained from an approximation valid for $\theta = {\cal O}
(N_g^{-1/3})$, has to be modified to agree with the boundary
condition $\psi(\theta=0)=0$. Indeed, when $\theta = {\cal O}
(N_g^{-1})$, we are in the same situation as in the boundary layer
$(I)$: the derivatives with respect to $\theta$ are of order
$\epsilon = {\cal O} (N_g^{-1})$, and the boundary layer equation is
eq.  (\ref{layerI}). The range of validity of the boundary layer
approximation $\psi^{III}_{in}$ is therefore
\begin{equation}
\psi^{III}_{in} = N_g^{-1/3}\phi^{(1)}(\xi,\tau), \;\;\;
N_g^{-1}\ll\theta\ll1.
\end{equation}

{\bf (IV).} A new boundary layer is needed for $\theta = {\cal
O}(N_g^{-1})$, in which the approximate transport equation is eq.
(\ref{layerI}). The solution of that equation satisfying the boundary
condition at $\theta=0$ is
\begin{equation}
\psi^{IV}_{in} \simeq C(\xi) \left(1-e^{-N_g\theta}\right).
\end{equation}
The arbitrary function $C(\xi)$ is obtained by imposing matching with
the approximate solution in the previous boundary layer
$\psi^{(III)}_{in}$. We obtain
\begin{equation}
\psi^{IV}_{in}(r,\theta) \simeq N_g^{-1/3}\phi^{(1)}(\xi,\tau=0)
\left(1-e^{-N_g\theta}\right), \;\;\; \theta\ll N_g^{-1/3}.
\end{equation}

This function gives the correction to the probability flux,
originated from the presence of the fixed particle, $\rho^{(1)}$:
\begin{equation}
\rho^{(1)}(r) \equiv \left.\frac{\partial\psi}{\partial z}\right|_{z=0}
\simeq N_g^{2/3}\phi^{(1)}(\xi,\tau=0), \;\,\;\;\; \xi=(r-1)N_g^{2/3}.
\end{equation}

Therefore, the numerical solution of eq. (\ref{layerIII}) specifies
the final probability distribution. One important conclusion is that,
for large values of $N_g$, $\rho^{(1)}$ scales to a function of the
form $N_g^{2/3}\phi^{(1)}[(r-1)N_g^{2/3}]$. It is easy to verify,
using eq. (\ref{normal}), that this distribution is normalized,
\begin{equation}
\int_1^{\infty}\rho^{(1)}(r) dr =
\int_0^{\infty}\phi^{(1)}(\xi,\tau=0)d\xi=1.
\end{equation}

\begin{figure}
\caption{Illustration of the governing equation for the non-uniform
deposition}
\label{gap}
\end{figure}

\begin{figure}
\caption{Probability density of adsorption in presence of a
pre-adsorbed particle as a function of distance for $N_g=10$. Solid
line: multipolar solution of transport equation (25) including 5
terms. Dots: Brownian dynamic simulations.}
\label{part1}
\end{figure}

\begin{figure}
\caption{Probability density of adsorption in presence of a
pre-adsorbed particle as a function of distance for $N_g=60$. Solid
line: exponential distribution (35). Dots: Brownian dynamic
simulations.}
\label{part1exp}
\end{figure}

\begin{figure}
\caption{Logarithmic of the scaled probability density of adsorption
in the presence of a pre-adsorbed particle. Simulation results for
$N_g=60$ (crosses) and $N_g=300$ (triangles). Continuous line:
Numerical solution of boundary layer equation.}
\label{bl}
\end{figure}

\begin{figure}
\caption{Comparison between the saturation coverage $\theta_\infty$
as a function of $R^*$ obtained with several methods.
(a) Dots: Brownian dynamic simulation.
(b) Solid line: numerical solution of the integral equation (7) using
the multipolar solution (25) truncated to five for $R^*<1.5$ and the
exponential fitting (35) for $R^* > 1.5$.
(c) Dashed line: asymptotic expression (39).
(d) Short dashed line: continuation of the numeric solution with
exponential fitting (35) for $R^*<1.5$.}
\label{recPe} 
\end{figure}

\begin{figure}
\caption{Probability density of adsorption in a line of length $40R$
in the presence of three adsorbed particles, at positions
$x_1=13R,\;x_2=20R,\;x_3=22.1R$.
(a) Dots: Brownian dynamic simulations.
(b) Solid line: superposition of the two one-particle solutions
corresponding to the particles limiting each gap.
(c)Dashed line: Superposition of the three one-particle solutions.}
\label{3part}
\end{figure}

\begin{figure}
\caption{Radial distribution function $g(r)$ for $R^*=1$. Dots:
Brownian dynamic simulations. Solid line: Calculation of $g(r)$ for
$1 \le r \le 2$ using the numerical solution of the integral equation
(7) for $n(h)$.}
\label{satgr1}
\end{figure}

\begin{figure}
\caption{Simulation results for $g(r)$ corresponding to $R^*=1.968$
($N_g=30$), squares and $R^*=3.310$ ($N_g=240$) displaced 4 units,
crosses. Solid line: Calculation of $g(r)$ for $1 \le r \le 2$ and
$N_g=30$ using the numerical solution of the integral equation (7).
Dashed line: the same for $N_g=240$.}
\label{satgr2}
\end{figure}


\begin{references}
\bibitem{ADSORPTION} Z. Adamczyk, B. Siwek, M. Zembala and P.
Belouschek, Adv. Colloid Interface Sci. {\bf 48}, 151 (1994).

\bibitem{FEDER} J. Feder and I. Giaever, J. Coll. Interface Sci. {\bf
78}, 144 (1980).

\bibitem{RAMSDEN} J. J. Ramsden, Phys. Rev. Lett. {\bf 71}, 295
(1993).

\bibitem{ONODA} G. Y. Onoda and E. G. Liniger, Phys. Rev. A {\bf 33},
715 (1986).

\bibitem{SCHAAF} P. Wojtaszczyk, P. Schaaf, B. Senger, M. Zembala and
J.-C. Voegel, J. Chem. Phys. {\bf 99}, 7198 (1993).

\bibitem{EVANS} J. W. Evans, Rev. Mod. Phys. {\bf 65}, 1281 (1993).

\bibitem{RSA} A. R\`enyi, Sel. Trans. Math. Stat. Prob. {\bf 4}, 205
(1963); E. R. Cohen and H. Reiss, J. Chem. Phys. {\bf 38}, 680
(1963); B. Widom, J. Chem. Phys. {\bf 44}, 2888 (1966); J. J.
Gonzalez, P. C. Hemmer and J. S. H{\o }ye, Chem. Phys. {\bf 3}, 288
(1974).

\bibitem{RSAsint} Y. Pomeau, J. Phys. A {\bf 13}, L193 (1980); R. H.
Swendsen, Phys. Rev. A {\bf 24}, 504 (1981).

\bibitem{RSAcomp} E. L. Hinrichsen, J. Feder and T. J. J{\o }ssang,
J. Stat. Phys. {\bf 44}, 793 (1986).

\bibitem{RSAg(r)} B. Bonnier, D. Boyer and P. Viot, J. Phys. A {27},
3671 (1994).

\bibitem{RSAapr} G. Tarjus, P. Schaaf and J. Talbot, J. Chem. Phys.
{\bf 93}, 8352 (1990); D. Boyer, G. Tarjus, P. Viot and J. Talbot, J.
Chem. Phys. {\bf 103}, 1607 (1995).

\bibitem{Virial} G. Tarjus, P. Schaaf and J. Talbot, J. Stat. Phys.
{\bf 63}, 167 (1991).

\bibitem{COLLOIDS} T. G. M. van de Ven, {\em Colloidal
Hydrodynamics}. Academic Press, New York (1989).

\bibitem{BALLISTIC} R. Jullien and P. Meakin, J. Phys. A {\bf 25},
L189 (1992); A. P. Thompson and E. D. Glandt, Phys. Rev. A {\bf 46},
4639 (1992); H. S. Choi, J. Talbot, G. Tarjus and P. Viot, J. Chem.
Phys. {\bf 99}, 9296 (1993).

\bibitem{BALexact} J. Talbot and S. M. Ricci, Phys. Rev. Lett. {\bf
68}, 958 (1992).

\bibitem{BALgen} P. Viot, G. Tarjus and J. Talbot, Phys. Rev. E {\bf
48}, 480 (1993).

\bibitem{DRSA} P. Schaaf, A. Johner and J. Talbot, Phys. Rev. Lett.
{\bf 66}, 1603 (1991); B. Senger, J.-C. Voegel, P. Schaaf, A. Johner,
A. Schmitt and J. Talbot, Phys. Rev. A {\bf 44}, 6926 (1991); B.
Senger, P. Schaaf, J.-C. Voegel, A. Johner, A. Schmitt and J. Talbot,
J. Chem. Phys. {\bf 97}, 3813 (1992); B. Senger, J. Talbot, P.
Schaaf, A. Schmitt and J.-C. Voegel, Europhys. Lett. {\bf 21}, 135
(1993).

\bibitem{ERMAK} D. L. Ermak and J. A. McCammon, J. Chem. Phys. {\bf
69}, 1352 (1978).

\bibitem{DRSAB} F.J. Bafaluy, H.S. Choi, B. Senger and J. Talbot,
Phys. Rev E.{\bf 51}, 5985 (1995).

\bibitem{DRSAG} B. Senger, J. Bafaluy, P. Schaaf, A. Schmitt and
J.-C. Voegel, Proc. Natl. Acad. Sci. USA {\bf 89}, 9449 (1992); B.
Senger, R. Ezzedine, J. Bafaluy, P. Schaaf, F. J. G. Cuisinier and
J.-C. Voegel, J. theor. Biol. {\bf 163}, 457 (1993);

\bibitem{Lleiescala} R. Ezzeddine, P.Schaaf, J.C. Voegel, B. Senger,
Phys. Rev E {\bf 51}, 6286 (1995).

\bibitem{HYDRO} J. Bafaluy, B. Senger, J.-C. Voegel and P. Schaaf,
Phys. Rev. Lett. {\bf 70}, 623 (1993).

\bibitem{BDHYDRO} I. Pagonabarraga and J.M. Rub\'{\i}, Phys. Rev.
Lett. {\bf 73}, 114 (1994).

\bibitem{Futur} J. Faraudo, J. Bafaluy (in preparation)

\bibitem{RAED} R. Ezzedine, P. Schaaf, J.-C. Voegel and B. Senger,
Phys. Rev. E {\bf 53}, 2473 (1996).

\bibitem{HANSEN} J.P. Hansen, I.R. McDonald {\em Theory of simple
liquids}, Academic Press, London (1990).

\bibitem{experim} P. Schaaf, P. Wojtaszczyk, E. K. Mann, B. Senger,
J.-C. Voegel and D. Bedeaux, J. Chem. Phys. {\bf 102}, 5077 (1995).

\bibitem{LAVENDA} B. Lavenda, {\em Statistical Physics: a
probabilistic approach}, J. Wiley \& sons, New York (1991).

\bibitem{ABRAMOWITZ} M. Abramowitz, I. Stegun, {\em Handbook of
Mathematical Functions}, New York: Dover Publications (1970).

\bibitem{KEVORKIAN} J. Kevorkian and J. D. Cole, {\em Perturbation
Methods in Applied Mathematics}, Springer, New York (1980).
\end{references}
\end{document}